\documentstyle[twocolumn,prl,aps]{revtex}

\begin{document}
\title{Quantum localisation observables and accelerated frames}
\author{Marc-Thierry Jaekel$^a$ and Serge Reynaud$^b$}
\address{$(a)$ Laboratoire de Physique Th\'eorique de l'ENS
\thanks{Laboratoire du CNRS associ\'e \`a
l'Ecole Normale Su\-p\'e\-rieu\-re et \`a l'Uni\-ver\-si\-t\'e Paris-Sud},
24 rue Lhomond, F75231 Paris Cedex 05 France\\
$(b)$ Laboratoire Kastler Brossel
\thanks{Laboratoire de l'Ecole Normale Su\-p\'e\-rieu\-re et de
l'Uni\-ver\-si\-t\'e Pierre et Marie Curie asso\-ci\'e au CNRS},
UPMC case 74, 4 place Jussieu, F75252 Paris Cedex 05 France}
\date{June 1998}
\maketitle

\begin{abstract}
We define quantum observables associated with Einstein localisation in
space-time. These observables are built on Poincar\'{e} and dilatation
generators. Their commutators are given by spin observables defined from the
same symmetry generators. Their shifts under transformations to uniformly
accelerated frames are evaluated through algebraic computations in conformal
algebra. Spin number is found to vary under such transformations with a
variation involving further observables introduced as irreducible quadrupole
momenta. Quadrupole observables may be dealt with as non commutative
polarisations which allow one to define step operators increasing or decreasing
the spin number by unity.
\end{abstract}

\section{Introduction}

In quantum field theory as well as in classical physics, space-time
parameters are introduced {\it a priori}, i.e. before the definition of any
other fundamental physical notions. It should however be obvious that
space-time is itself a physical notion which has to be confronted with the
necessity of realising time and space standards and of comparing time and
space intervals. These realisations and comparisons have to rely on physical
systems and, ultimately, on the laws of physics. It was clearly demonstrated
by Einstein \cite{Einstein05} that such a physical conception of space-time
has drastic consequences gathered under the general denomination of
relativistic effects. This conception as well as its relativistic
consequences play nowadays a key role in the metrological realisation of
space-time units \cite{IEEE91} as well as in the definition of reference
systems \cite{WolfPetit,Guinot97}.

A first step in a constructive approach to space-time is the definition of
localisation procedures. In order to describe physical phenomena localised
in space and time, it is indeed necessary to have the ability to define
event times at different locations in space and, then, to establish
relations between these event times. These two requirements may be
respectively termed as time definition and time transfer or, alternatively,
as clock realisation and clock synchronisation. Introductory presentations
of relativity, as well as now existing practical localisation systems such
as the Global Positioning System \cite{GPS91}, are based on time transfers
between remote observers exchanging electromagnetic signals. The
electromagnetic field is thus used as a support to encode a time reference
used for comparing clock indications. A localisation procedure may then be
built as the result of several time transfers. These constructions clearly
rely on the existence of a universal field propagation velocity, the
velocity of light $c$. In other words, the relativistic notion of space-time
is ultimately based upon the symmetries of field propagation.

In particular, faithfulness of synchronisation procedures requires that the
references be defined from observables preserved by propagation.
Localisation in space-time should therefore be built on the conserved
quantities associated with symmetries of field propagation. On another hand,
these symmetries constitute the fundamental expression of relativistic laws
determining the effects of space-time transformations between moving frames.
They also play a primary role in metrology. Translation symmetry allows one to
transport metrological standards from one place to another. Lorentz symmetry
permits one to use standards in different inertial frames and to derive a
length unit from the time unit. These discussions look familiar since the
invariance of Maxwell equations under Poincar\'{e} transformations played a
prime role in Einstein's introduction of relativitistic theories. The role
played by dilatation is less often discussed although the invariance of
Maxwell equations under dilatations has been known for a long time \cite
{Pais82}. Dilatations are naturally involved in comparisons of lengths or
durations with different scales. Appropriate behaviours under dilatation
have in fact to be considered as symmetry requirements for the problem of
unit definition.

Furthermore, Maxwell equations are also invariant under the group of
conformal coordinate transformations \cite{Bateman09,Cunningham09}. This
invariance may be understood as manifesting the insensitivity of light
propagation to a conformal metric factor \cite{MashhoonG80}, that is also to
a change of space-time scale. The conformal coordinate transformations not
only include transformations from inertial frames to other inertial frames,
but also transformations to accelerated frames \cite{FultonRW62}. Conformal
symmetry should therefore allow one to derive the shifts of observables
under such transformations to accelerated frames or, in other words, to
obtain redshifts \cite{Einstein07} from invariance properties rather than
from covariance properties \cite{Norton93}.

Relativitistic concepts were introduced in the context of classical
relativity where observables are represented by real numbers which can, in
principle, be determined with arbitrary precision. They have to remain
pertinent in a quantum context where observables possess quantum
fluctuations and can no longer be given a classical representation. This
raises novel challenges that we may characterise as the definition of a
quantum relativity. Possible ways to take up these challenges are clearly
indicated by the previous arguments. Localisation in space-time has to be
described in terms of quantum observables related to the symmetry generators of
field propagation.

Preliminary results have already been obtained by following this approach
\cite{PRL96,PLA96,EPL97}.
The algebraic technique developed may be characterised as an
embedding of the symmetry algebra in the algebra of quantum observables.
 All properties can be derived from the conformal
algebra, that is the set of commutators between the symmetry generators.
The generators contained in quantum algebra are used to define localisation
observables, and their commutators
to describe their quantum commutation relations as well
as their relativistic shifts under frame transformations.
As a consequence, localisation observables can be defined in a quantum
framework while being fully consistent with relativistic requirements.
In the present paper, we will give a complete
characterisation of quantum observables associated with the problem of
localisation in space-time and of their shifts under transformations to
accelerated frames.

An important output of this quantum algebraic technique is that the shifts
do not keep their form unaltered when transfered from classical to quantum
relativity. In particular, mass and spin number, defined as Casimir
invariants of the Poincar\'{e} algebra, will be shown to vary under
transformations to accelerated frames. This is not too surprising since mass
and spin number defined in this manner are quantum observables which cannot
be reduced {\it a priori} to classical numbers. The shift of mass will be
described by a conformal factor depending on position as expected from the
equivalence principle. Although it is invariant under Poincar\'{e}
transformations as well as dilatations, the spin number will be found to
vary under conformal transformations to accelerated frames. Its variation
will be shown to involve further observables representing irreducible
quadrupole momenta of the quantum distribution of energy-momentum density.

Throughout the main body of the paper, we will consider the generic case of
arbitrary field states. To make the connections of our approach with
standard quantum field theory more explicit, we will however study the
specific cases of $1-$photon and $2-$photon states in appendices \ref
{appConfInv}-\ref{appLocalisation}. We will give explicit expressions of the
time reference transfered between remote observers and of the space-time
localisation observables. We will also discuss a geometrical interpretation
of localisation observables.

\section{Poincar\'{e} and dilatation algebras}

As a first step, we recall the basic properties of symmetry algebras as they
are known for Poincar\'{e} and dilatation algebras, and how they are
embedded in the quantum algebra of observables.

Poincar\'{e} transformations are described by $10$ generators, namely the $4$
components $P_{\mu }$ representing translations and the $6$ independent
components of the antisymmetric tensor $J_{\mu \nu }$ ($J_{\nu \mu }=-J_{\mu
\nu }$) representing rotations and Lorentz boosts. All the symmetry
properties associated with special relativity are described by the
Poincar\'{e} algebra, that is the set of commutators between these
generators
\begin{eqnarray}
&&\left( P_{\mu },P_{\nu }\right) =0\qquad \left( J_{\mu \nu },P_{\rho
}\right) =\eta _{\nu \rho }P_{\mu }-\eta _{\mu \rho }P_{\nu }  \nonumber \\
&&\left( J_{\mu \nu },J_{\rho \sigma }\right) =\eta _{\nu \rho }J_{\mu
\sigma }+\eta _{\mu \sigma }J_{\nu \rho }-\eta _{\mu \rho }J_{\nu \sigma
}-\eta _{\nu \sigma }J_{\mu \rho }  \label{PoincareAlg}
\end{eqnarray}
In quantum field theory, the symmetry generators are identified with the
conserved quantities derived from Noether's theorem \cite{ItzyksonZ85}. For
completeness, we recall the relations between the
symmetry generators and quantum
fields in appendix \ref{appConfInv}. The generators $P_{\mu }$ are the
energy-momentum operators whereas the generators $J_{\mu \nu }$ represent
angular momentum components in the four-dimensional space-time. $\eta _{\mu
\nu }$ is the Minkowski tensor
\begin{equation}
\eta _{\mu \nu }\equiv {\rm diag}\left( 1,-1,-1,-1\right)
\end{equation}
used throughout the paper to raise or lower tensor indices and to express
scalar products. We also denote $\eta _{\mu }^{\rho }$ the Kronecker symbol.
Commutators of observables are written as the usual quantum commutators
divided by $i\hbar $
\begin{equation}
\left( A,B\right) \equiv \frac{1}{i\hbar }\left[ A,B\right] \equiv \frac{1}{%
i\hbar }\left( AB-BA\right)
\end{equation}
They obey the Jacobi identity
\begin{equation}
\left( \left( A,B\right) ,C\right) =\left( A,\left( B,C\right) \right)
-\left( B,\left( A,C\right) \right)   \label{Jacobi}
\end{equation}

The relations embedded in Poincar\'{e} algebra mean that the generators
belong to the algebra of quantum observables with characteristic commutation
relations (\ref{PoincareAlg}). At the same time, they entail that the
generators are relativistic observables which are shifted under frame
transformations according to the same relations (\ref{PoincareAlg}). Since
the generators are quantum observables, we will have to take care of their
non-commutativity. To this aim, we will use a symmetrised product
which has to be manipulated with care since it is not associative
\begin{eqnarray}
A\cdot B &\equiv &\frac{1}{2}\left( AB+BA\right)  \nonumber \\
A\cdot \left( B\cdot C\right) -\left( A\cdot B\right) \cdot C &=&\frac{\hbar
^{2}}{4}\left( B,\left( A,C\right) \right)
\end{eqnarray}
Occasionally, we will use the dot symbol to represent at once a symmetrised
scalar product of two vectors
\begin{equation}
A\cdot B\equiv A^{\rho }\cdot B_{\rho }
\end{equation}

Poincar\'{e} algebra has two Casimir invariants, the mass and the squared
spin. The squared mass $P^{2}$ is defined for an arbitrary physical state as
the norm of energy-momentum vector and is invariant under all Poincar\'{e}
transformations \cite{Einstein0506}
\begin{equation}
\left( P_{\mu },P^{2}\right) =\left( J_{\mu \nu },P^{2}\right) =0
\label{PP2}
\end{equation}
Spin observables are introduced in a relativistic framework through the
Pauli-Lubanski vector \cite{ItzyksonZ85}
\begin{eqnarray}
W^{\mu } &\equiv &-\frac{1}{2}\epsilon ^{\mu \nu \rho \sigma }J_{\nu \rho
}P_{\sigma }  \nonumber \\
\left( P_{\mu },W_{\rho }\right)  &=&0\qquad \left( J_{\mu \nu },W_{\rho
}\right) =\eta _{\nu \rho }W_{\mu }-\eta _{\mu \rho }W_{\nu }  \label{PW}
\end{eqnarray}
$\epsilon _{\mu \nu \lambda \rho }$ is the completely antisymmetric Lorentz
tensor
\begin{eqnarray}
\epsilon _{{\rm 0123}} &=&-\epsilon ^{{\rm 0123}}=+1  \nonumber \\
\epsilon _{\mu \nu \rho \sigma } &=&-\epsilon _{\mu \nu \sigma \rho
}=-\epsilon _{\mu \rho \nu \sigma }=-\epsilon _{\nu \mu \rho \sigma }
\end{eqnarray}
The commutators between components of the spin vector may be written in
terms of a spin tensor
\begin{equation}
\left( W_{\mu },W_{\nu }\right) =P^{2}S_{\mu \nu }=\epsilon _{\mu \nu \rho
\sigma }W^{\rho }P^{\sigma }  \label{defSt}
\end{equation}
The case of a vanishing mass raises a problem for extracting the spin
tensor. This case will be discussed later on. Spin observables commute with
momentum and they are transverse with respect to momentum
\begin{equation}
P^{\mu }S_{\mu \nu }=P_{\mu }W^{\mu }=0  \label{trans}
\end{equation}

Since $W^{\mu }$ is a Lorentz vector, its squared modulus is a Lorentz
scalar that we can write under its standard form in terms of a spin number $s
$ taking integer or half-integer values
\begin{equation}
S^{2}=\frac{W^{2}}{P^{2}}=\frac{1}{2}S_{\mu \nu }S^{\nu \mu }=-\hbar
^{2}s\left( s+1\right)
\end{equation}
The negative sign in the relation between $S^{2}$ and $s\left( s+1\right) $
corresponds to the fact that spin is a space-like vector. For the sake of
simplicity, we have set the velocity of light to unity. However, we keep the
Planck constant $\hbar $ as the characteristic scale of quantum effects.

To describe the dilatation symmetry, we enlarge the Poincar\'{e} algebra by
a dilatation generator $D$ and further commutation relations
\begin{equation}
\left( D,P_{\mu }\right) =P_{\mu }\qquad \left( D,J_{\mu \nu }\right) =0
\label{DAlg}
\end{equation}
Generally speaking, commutation relations with $D$ may be thought as
defining the conformal weight of observables. This weight vanishes for $%
J_{\mu \nu }$ but not for $P_{\mu }$. The spin number is a Poincar\'{e}
invariant with a null conformal weight
\begin{equation}
\left( P_{\mu },s\right) =\left( J_{\mu \nu },s\right) =\left( D,s\right) =0
\label{Ps}
\end{equation}

\section{Localisation observables}

As discussed in the introduction, we intend to define the space-time
position of an event from the symmetry generators. Precisely, we will use
the Poincar\'{e} and dilatation generators which are symmetry generators for
the electromagnetic field used in Einstein synchronisation or localisation
procedures.

We first recall results which have already been derived by using a simple
two-dimensional model. In a synchronisation procedure, a time reference is
transfered between two remote observers through the exchange of a light
pulse. Classically, this reference is the value of the light cone variable
preserved under field propagation. In quantum theory, a
similar reference observable may be defined from the
translation and dilatation generators
corresponding to the field propagating in this single direction
\cite{PRL96}.
Since this observable is a transfer variable, i.e. a light
cone variable, its space-time components are only defined in the
direction transverse with respect to the line of sight. When two transfer
procedures are performed along counterpropagating directions, two light cone
variables may be exchanged between
remote observers, allowing them to obtain the position of the other one in
space and time. Basically, this localisation procedure amounts to associate
a position in space-time with the coincidence event corresponding to the
intersection of two light pulses \cite{PLA96}. Clearly, this description
heavily relies on a specific feature of two-dimensional field theories,
namely the existence of an {\it a priori} decomposition of fields in
counterpropagating directions.

In four-dimensional space-time in contrast, such a natural decomposition is
not available. Furthermore, light rays have an intrinsic transverse
extension due to diffraction and two light rays do not necessarily cross
each other. The description of synchronisation and localisation procedures
may nonetheless be given following the same ideas. This can be illustrated
by using specific electromagnetic field states, namely $1-$photon states for
synchronisation and $2-$photon states for localisation as analysed in detail
in appendices \ref{appSynchronisation}-\ref{appLocalisation}.
As a result, the total Poincar\'e and dilatation generators of the $2-$photon
field are sufficient to determine the position of the coincidence event.
In the main body of the paper, we show
how a position in space-time can be defined for an arbitrary field state
from the generators of Poincar\'{e} and dilatation symmetries.

To build up this definition, we first write the angular momentum components $%
J_{\mu \nu }$
of the total field
as sums of orbital contributions having their usual form in
terms of momenta and positions and of spin contributions (\ref{defSt})
\begin{equation}
J_{\mu \nu }=P_{\mu }\cdot X_{\nu }-P_{\nu }\cdot X_{\mu }+S_{\mu \nu }
\label{J_X}
\end{equation}
These relations alone are not sufficient to determine the expression of
position observables since they do not fix their longitudinal part aligned
along momentum. A simple assumption to fix this longitudinal part is to
identify the generator $D$ as the scalar product of momentum and position
vectors \cite{PLA96,EPL97}
\begin{equation}
D=P\cdot X  \label{D_X}
\end{equation}
Relations (\ref{J_X}-\ref{D_X}) lead to the following relation between
position observables, Poincar\'{e} and dilatation generators
\begin{equation}
P^{2}\cdot X_{\mu }=P_{\mu }\cdot D+P^{\rho }\cdot J_{\rho \mu }
\end{equation}
The extraction of $X_{\mu }$ from this relation requires a non vanishing
mass, as the extraction of the spin tensor from (\ref{defSt}). We have thus
to face two different situations. When mass associated with the field state
vanishes, localisation observables cannot be completely defined. This
corresponds in fact to a synchronisation case and occurs in particular when
the field contains a single photon. We may then define a transfer observable
which is only defined transversely to the transfer propagation and can be
exchanged by two remote observers (see appendix \ref{appSynchronisation}).
This observable clearly generalises the transfer variable more easily
defined in a two-dimensional quantum field theory \cite{PRL96}. The latter
can be considered as extending the Newton-Wigner definition of positions
\cite{NewtonW49} in a Lorentz covariant manner \cite{BJP95}.

When the field contains photons propagating in at least two different
directions, the mass no longer vanishes and the field state provides us with
a quantum definition of space-time localisation observables
\begin{equation}
X_{\mu }=\frac{P_{\mu }}{P^{2}}\cdot D+\frac{P^{\rho }}{P^{2}}\cdot J_{\rho
\mu }  \label{defX}
\end{equation}
The algebraic properties of these observables follow from the symmetry
algebras (\ref{PoincareAlg},\ref{DAlg}). The localisation observables are
shifted under translations, dilatation and rotations exactly as ordinary
coordinate parameters are shifted under the corresponding transformations in
classical relativity
\begin{eqnarray}
&&\left( P_{\mu },X_{\nu }\right) =-\eta _{\mu \nu }\qquad \left( D,X_{\mu
}\right) =-X_{\mu }  \nonumber \\
&&\left( J_{\mu \nu },X_{\rho }\right) =\eta _{\nu \rho }X_{\mu }-\eta _{\mu
\rho }X_{\nu }  \label{PX}
\end{eqnarray}
The shifts under translations mean that position observables $X_{\mu }$ are
canonically conjugate to momenta. The commutators of different components of
positions (\ref{defX}) may also be deduced
\begin{equation}
P^{2}\cdot \left( X_{\mu },X_{\nu }\right) =S_{\mu \nu }  \label{XX}
\end{equation}
These commutators do not vanish in the general case of a non vanishing spin.
This constitutes a manifestation in the present formalism of the known
problem of localisability in the presence of spin \cite{Pryce48,Fleming65}.
This is also a clear evidence that concepts originating from classical
conceptions of space-time have to be modified in a quantum and relativistic
theoretical framework.

The observable $X_{\mu }$ is a position in time for $\mu =0$ and a position
in space for $\mu =1,2,3$. Relations (\ref{PX}) thus mean that a time
observable has been defined which is conjugate to energy in the same manner
as space observables are conjugate to spatial momenta. An energy-time
commutation relation exists which effectively asserts that the fourth
Heisenberg inequality constrains quantum fluctuations of time and energy
\cite{Jammer74}. The observables $X_{\mu }$ are built from conserved
quantities and, consequently, they do not evolve due to field propagation.
Hence, they are conceptually different from coordinate parameters used for
describing evolution. In particular, the time observable $X_{0}$ represents
a date, i.e. the position of an event in time. As a date, it does not evolve
and cannot be confused with the affine parameter used to write equations of
motion.

We have thus defined positions in space and time in a Lorentz covariant
manner. Furthermore, we have described their transformations under
Poincar\'{e} and dilatation generators by Lorentz covariant formulas (\ref
{PX}). This is an answer to the long standing riddle raised by the relation
between time and space definitions in quantum theory on one hand and
relativistic effects associated with Lorentz transformations on the other
hand \cite{Schrodinger30,Rovelli91}.

It is worth stressing that position observables cannot be defined when the
mass associated with the field state vanishes. To define positions (\ref{defX}%
), at least $2$ photons propagating in different directions are needed. This
means in particular that the domain of definition of localisation
observables does not cover the space of all field states since it excludes
vacuum and $1-$photon states. Hence, position operators are not
self-adjoint. This has often been considered as an objection against the
very possibility of giving a quantum definition of phase or of time \cite
{Pauli80}. However position operators are
examples of hermitic
but not self-adjoint observables \cite{BogolubovLT}
which have been repeatedly shown to allow for a rigorously
consistent treatment, as exemplified by the formalism of positive operator
valued measures \cite{LevyLeblond76,BuschGL95}. In the present approach,
this problem is dealt with by a calculus operating in the algebra of
observables defined as the enveloping division ring built on symmetry
algebra. This quantum algebraic calculus is rigorously defined as soon as
divisions by $P^{2}$ are carefully dealt with which, of course, restricts
the domain of validity of some relations to states where $P^{2}$ differs
from zero \cite{Toller98}.

In the specific case of $2-$photon states, a geometric interpretation of the
positions $X_{\mu }$ may be given. It is analysed in detail in appendix \ref
{appLocalisation}.

\section{Transformations to accelerated frames}

As already discussed in the introduction, invariance of electromagnetism
under the group of conformal transformations will allow us to deal with
tranformations to accelerated frames.

To this aim, we introduce the whole set of conformal generators which
contains the $11$ already discussed generators, $10$ for Poincar\'{e}
transformations and $1$ for dilatations, and $4$ additional ones $C_{\mu }$
representing conformal transformations to accelerated frames. These
generators may be defined as integrals of the electromagnetic stress tensor
in the usual manner \cite{BesselHagen21}. Conformal invariance can be
rigorously established for quantum electromagnetic fields \cite{BinegarFH83}
. We will consider here that all generators vanish in vacuum, in consistency
with conformal invariance of electromagnetic vacuum \cite{QSO95}. More
generally, the definition of photon number is conformally invariant \cite
{Gross64}. More precise statements of these properties are given in appendix
\ref{appConfInv}.

Conformal algebra contains commutators (\ref{PoincareAlg},\ref{DAlg})
complemented by the following ones
\begin{eqnarray}
&&\left( C_{\mu },C_{\nu }\right) =0\qquad \left( D,C_{\mu }\right) =-C_{\mu
}  \nonumber \\
&&\left( P_{\mu },C_{\nu }\right) =-2\eta _{\mu \nu }D-2J_{\mu \nu }
\nonumber \\
&&\left( J_{\mu \nu },C_{\rho }\right) =\eta _{\nu \rho }C_{\mu }-\eta _{\mu
\rho }C_{\nu }  \label{ConfAlg}
\end{eqnarray}
The four new generators are commuting components of a vector and they have a
conformal weight opposite to that of momenta. Commutators in the second line
of (\ref{ConfAlg}) describe the shifts of energy-momentum under
transformations to accelerated frames and will be interpreted in the
following as quantum expressions of the Einstein redshift law.

To discuss the shifts of observables under transformations to accelerated
frames, we introduce the definition $\Delta _{a}$ for such a generic
transformation
\begin{equation}
\Delta _{a}=\frac{a^{\mu }}{2}C_{\mu }  \label{defDelta}
\end{equation}
where the classical numbers $a^{\mu }$ represent accelerations along the
four space-time directions. As a first example, we evaluate the redshift of
mass
\begin{equation}
\left( \Delta _{a},P^{2}\right) =2a^{\mu }P^{2}\cdot X_{\mu }  \label{CM}
\end{equation}
This relation could also be considered as defining quantum positions in
space-time. As a matter of fact, the potential energy of a mass in a
constant gravitational field is proportional to mass and to a gravitational
potential depending linearly on the position measured along the direction of
gravity. The equivalence between constant gravity and uniform acceleration
then implies to read the redshift of mass as a definition of position \cite
{EPL97}. Notice that this expression is valid for vanishing mass but gives
an unambiguous definition of position only for states corresponding to a non
vanishing mass. The mass shift (\ref{CM}) may also be read as a conformal
metric factor arising in transformations to accelerated frames and depending
on position observables as the classical metric factor depends on classical
position \cite{FoP97}.

To prevent any confusion, let us emphasise that the redshift of mass (\ref
{CM}) does not constitute a violation but rather a consequence of conformal
symmetry of electromagnetism. There is nothing paradoxical in this situation
which is familiar in relativistic theories. For instance, time is an
absolute of classical physics which is shown to vary in relativistic physics
as a consequence of the symmetry of electromagnetism under Lorentz
tranformations.

After the redshift of mass, we now write the redshift of momenta as
\begin{eqnarray}
\left( \Delta _{a},P_{\nu }\right)  &=&a_{\nu }D-a^{\mu }J_{\mu \nu }
\nonumber \\
&=&a_{\nu }P\cdot X-a^{\mu }P_{\mu }\cdot X_{\nu }+a^{\mu }X_{\mu }\cdot
P_{\nu }  \nonumber \\
&&-a^{\mu }S_{\mu \nu }  \label{CP}
\end{eqnarray}
where we have used (\ref{J_X},\ref{D_X},\ref{ConfAlg}). This quantum
redshift law differs from the classical one as a consequence of the spin
dependence. When the redshift of mass (\ref{CM}) is evaluated, the spin
dependence however disappears as a consequence of transversality relations (%
\ref{trans}). Notice that both redshift laws (\ref{CM}-\ref{CP}) have a
universal form dictated by conformal algebra, although the latter form
differs from the classical one.

One aim of the present paper is to derive the shifts of positions $X_{\mu }$
under transformations to accelerated frames. This derivation will require
further developments but we may already get some fruitful insights on the
universality of relativistic transformations. To this aim, we note that the
canonical commutators (\ref{PX}) are invariant under all frame
transformations and in particular under $\Delta_{a}$
\begin{equation}
\left( \Delta_a ,\left( P_{\mu },X_{\nu }\right) \right) =0
\end{equation}
Jacobi identity (\ref{Jacobi}) then leads to the following relation
\begin{equation}
\left( \left( \Delta_a ,X_{\nu }\right) ,P_{\mu }\right) =\left( \left(
\Delta_a
,P_{\mu }\right) ,X_{\nu }\right)   \label{CXP}
\end{equation}
Using (\ref{PX},\ref{CP}), the second expression is explicitly evaluated as
\begin{equation}
\left( \left( \Delta_a ,P_{\mu }\right) ,X_{\nu }\right) =-\eta _{\mu \nu
}a\cdot X-a_{\mu }X_{\nu }+a_{\nu }X_{\mu }  \label{CPX}
\end{equation}
These results entail that we already know the shift under a translation $%
\left( \left( \Delta_a ,X_{\nu }\right) ,P_{\mu }\right) $ of the shift of
position $\left( \Delta_a ,X_{\nu }\right) $ under transformations to
accelerated frames. Furthermore, this expression has a classical form which
generalises in a quantum framework the covariance rules of classical
relativity \cite{EPL97}. It will be used as a consistency test when the
complete expression for the shift of position $\left( \Delta_a ,X_{\nu
}\right) $ will be available.

Proceeding further, we notice that momentum, position and spin are
sufficient to build up a conformal algebra, that is a set of generators
satisfying commutators (\ref{ConfAlg}). Indeed, the following expression
provides a realisation of conformal generators as non linear functions of
Poincar\'{e} and dilatation generators
\begin{equation}
2D\cdot X_{\mu }-P_{\mu }\cdot X^{2}+2X^{\rho }\cdot S_{\rho \mu }-\frac{%
P_{\mu }}{P^{2}}S^{2}  \label{defCp}
\end{equation}
Precisely, commutation relations (\ref{ConfAlg}) are obeyed when $C_{\mu }$
is replaced by this expression. Consequently, the redshift laws (\ref{CM},%
\ref{CP}) are also unchanged with $C_{\mu }$ replaced by (\ref{defCp}). This
does not mean however that the generators $C_{\mu }$ which represent the
symmetry of field propagation under transformations to accelerated frames
may be reduced to the expression (\ref{defCp}). Such a reduction would imply
peculiar constraints on the field states which are not satisfied in general
\cite{PRL96,PLA96}.

\section{Quadrupole observables}

We now introduce quadrupole observables which are precisely defined from the
differences between $C_{\mu }$ and (\ref{defCp}). These observables are
further observables of interest for the problem of localisation. The
characterisation of this problem comes to an end with this new definition
since the shifts of quadrupoles may be written in terms of known observables
including quadrupoles.

To facilitate reading of forthcoming derivations, it is convenient to
introduce a mass operator defined as the square root of $P^{2}$
\begin{equation}
M=\sqrt{P^{2}}  \label{defM}
\end{equation}
This is a Lorentz scalar with a non null conformal weight
\begin{eqnarray}
\left( P_{\mu },M\right)  &=&\left( J_{\mu \nu },M\right) =0  \nonumber \\
\left( D,M\right)  &=&M  \label{PM}
\end{eqnarray}
It may then be used to bring the conformal weight of vectors to zero. In
particular, one may define weightless vectors from momentum and
Pauli-Lubanski vectors
\begin{equation}
V_{\mu }\equiv \frac{P_{\mu }}{M}\qquad S_{\mu }\equiv \frac{W_{\mu }}{M}
\end{equation}
The first one is a velocity vector and the second one a spin vector. Both
obey the following generic relations of invariance under translations and
dilatations, and rotation as a Lorentz vector
\begin{eqnarray}
&&\left( P_{\mu },A_{\rho }\right) =0\qquad \left( D,A_{\rho }\right) =0
\nonumber \\
&&\left( J_{\mu \nu },A_{\rho }\right) =\eta _{\nu \rho }A_{\mu }-\eta _{\mu
\rho }A_{\nu }  \label{PVector}
\end{eqnarray}
These properties allow us to derive the following commutation relations with
position and spin observables
\begin{eqnarray}
\left( X_{\mu },A_{\rho }\right)  &=&\eta _{\mu \rho }\frac{V\cdot A}{M}-%
\frac{A_{\mu }V_{\rho }}{M}  \nonumber \\
\left( S_{\mu },A_{\rho }\right)  &=&A_{\mu \rho }=\epsilon _{\mu \rho \nu
\sigma }A^{\nu }V^{\sigma }  \nonumber \\
\left( S_{\mu \nu },A_{\rho }\right)  &=&\left( \eta _{\nu \rho }-V_{\nu
}V_{\rho }\right) \left( A_{\mu }-\left\{ V\cdot A\right\} V_{\mu }\right)
\nonumber \\
&-&\left( \eta _{\mu \rho }-V_{\mu }V_{\rho }\right) \left( A_{\nu }-\left\{
V\cdot A\right\} V_{\nu }\right)   \nonumber \\
\left( S_{\mu }{}^{\rho },A_{\rho }\right)  &=&2\left( A_{\mu }-\left\{
V\cdot A\right\} V_{\mu }\right)   \nonumber \\
A^{\mu } &=&\left\{ V\cdot A\right\} V^{\mu }-\frac{1}{2}\epsilon ^{\mu \nu
\rho \sigma }A_{\nu \rho }V_{\sigma }  \label{XVector}
\end{eqnarray}
The order of $V$ and $A$ does not matter since they commute. We have
introduced a tensor representation $A_{\mu \rho }$ of the vector $A^{\mu }$.
The scalar $V\cdot A$ commutes with all observables built upon Poincar\'{e}
generators in particular with position and spin observables. When this
scalar vanishes, the vector is transverse with respect to momentum and it
therefore obeys simplified relations. In particular, $S_{\mu }$ is such a
transverse weightless vector obeying these equations.

We now come to a decomposition of the generators $C_{\mu }$ as sums of
already known contributions (\ref{defCp}) and of further ones
\begin{eqnarray}
C_{\mu } &=&2D\cdot X_{\mu }-P_{\mu }\cdot X^{2}+2X^{\rho }\cdot S_{\rho \mu
}-\frac{P_{\mu }}{P^{2}}S^{2}  \nonumber \\
&&+2\hbar \frac{Q_{\mu }}{M}  \label{defQ}
\end{eqnarray}
This separation is in fact analogous to equation (\ref{J_X}) where the
angular momentum $J_{\mu \nu }$ was written as the sum of an orbital
contribution built on momenta and positions and of further spin observables
which may be thought of as internal angular momenta. In (\ref{defQ}), the
first line represents external contributions to $C_{\mu }$ built on momenta,
position and spin observables while the second line represents internal
contributions describing the dispersion of momentum distribution.
Observables $Q_{\mu }$ will be called quadrupole momenta in the following.
They are defined so that they scale as the Planck constant $\hbar $ like the
spin observables. They obey equations (\ref{PVector}), so that their
commutation relations with position and spin operators are given by (\ref
{XVector}).

We are now able to write the shifts of localisation observables under
transformations to accelerated frames
\begin{eqnarray}
\left( \Delta _{a},X_{\nu }\right) &=&\frac{a_{\nu }}{2}X^{2}-a^{\mu }X_{\mu
}\cdot X_{\nu }  \nonumber \\
&+&a^{\mu }\frac{S_{\mu }\cdot S_{\nu }}{M^{2}}-\frac{a_{\nu }}{2}\frac{S^{2}%
}{M^{2}}  \nonumber \\
&+&a^{\mu }\frac{\hbar }{M^{2}}\left\{ \eta _{\mu \nu }V\cdot Q-V_{\mu
}\cdot Q_{\nu }-V_{\nu }\cdot Q_{\mu }\right\}  \label{CX}
\end{eqnarray}
The first two lines correspond to the contribution of the external part (\ref
{defCp}) of $C_{\mu }$. The first line contains terms proportional to
positions which coincide with the shifts expected from classical relativity.
The second line contains terms depending on spin which thus appear as
quantum corrections to classical expressions. Finally, the third line
contains quadrupole corrections. All quantum corrections, that is spin and
quadrupole corrections, scale as $\frac{\hbar ^{2}}{M^{2}}$ and have to be
compared with classical terms scaling as $X^{2}$. Let us note that quantum
terms (second and third lines) commute with momenta operators, so that only
the classical terms (first line) contribute when the quantities $\left(
\left( \Delta _{a},X_{\nu }\right) ,P_{\mu }\right) $ are evaluated. In
other words, equations (\ref{CXP}-\ref{CPX}) are recovered from (\ref{CX}).

The shifts of spin observables may be written similarly
\begin{eqnarray}
\left( \Delta _{a},S_{\nu }\right)  &=&a_{\nu }X\cdot S-a^{\mu }S_{\mu
}\cdot X_{\nu }  \nonumber \\
&&+\frac{\hbar }{M}a^{\mu }Q_{\mu \nu }  \label{CS}
\end{eqnarray}
The first line contains classical looking terms while the second one
contains quadrupole corrections. The tensor $Q_{\mu \nu }$ is defined from
the vector $Q_{\mu }$ according to (\ref{XVector}). Here quadrupole terms
appear as corrections of order $\frac{\hbar }{M}$ with respect to the
standard terms. The classical terms are such that the squared spin $S^{2}$
and therefore the spin number $s$ are preserved. But this is not always the
case for the quadrupole corrections as shown by the following relation
\begin{eqnarray}
&&\left( C_{\mu },S^{2}\right) =4\hbar ^{2}\frac{R_{\mu }}{M}  \nonumber \\
&&R_{\mu }=Q_{\mu \nu }\cdot \frac{S^{\nu }}{\hbar }=\epsilon _{\mu \nu \rho
\sigma }Q^{\rho }\cdot \frac{S^{\nu }}{\hbar }V^{\sigma }  \label{defR}
\end{eqnarray}
The vector $R_{\mu }$ does not introduce new observables since it is defined
as a four-dimensional vectorial product of velocity, spin and quadrupole
vectors. It is orthogonal to velocity and spin vectors
\begin{equation}
V_{\mu }R^{\mu }=R^{\mu }V_{\mu }=S_{\mu }R^{\mu }=R^{\mu }S_{\mu }=0
\label{transR}
\end{equation}
Furthermore, it is invariant under translations and dilatations so that it
obeys relations (\ref{XVector}) with the simplification associated with
transversality.

The commutation relations of quadrupole components may also be obtained from
conformal algebra
\begin{equation}
\left( Q_{\mu },Q_{\nu }\right) =2\left\{ \frac{V\cdot Q}{\hbar }S_{\mu \nu
}+R_{\mu }\cdot V_{\nu }-R_{\nu }\cdot V_{\mu }\right\}  \label{QQ}
\end{equation}
As an important consequence, the shift of quadrupole observables under
transformations to accelerated frames may be written in terms of already
known localisation observables including quadrupoles
\begin{eqnarray}
\left( \Delta _{a},Q_{\nu }\right) &=&a_{\nu }X\cdot Q-a^{\mu }Q_{\mu }\cdot
X_{\nu }  \nonumber \\
&+&\frac{\hbar a^{\mu }}{M}\left\{ \frac{V\cdot Q}{\hbar }S_{\mu \nu
}+R_{\mu }\cdot V_{\nu }-R_{\nu }\cdot V_{\mu }\right\}  \label{CQ}
\end{eqnarray}
Hence, it will not be necessary to introduce further observables to obtain a
full characterisation of the shifts of localisation observables. Expressions
(\ref{CX},\ref{CS},\ref{CQ}) provide such a characterisation in the general
case of an arbitrary quantum state.

Relations (\ref{CM}) and (\ref{defR}) show that the two Casimir invariants
of Poincar\'{e} algebra are not invariant under conformal transformations to
accelerated frames. The Casimir invariants of conformal algebra can be
obtained by examining quantities already known to be invariant under
Poincar\'{e} and dilatation generators. There exist four non trivial
quantities of this kind, namely $S^{2}$, $Q^{2}$, $V\cdot Q$ and $S\cdot Q$.
Their shifts under transformations to accelerated frames are found to be
\begin{eqnarray}
&&\left( C_{\mu },\hbar V\cdot Q\right) =\left( C_{\mu },\frac{Q^{2}}{2}%
\right) =\left( C_{\mu },S^{2}\right)   \nonumber \\
&&\left( C_{\mu },S\cdot Q\right) =0  \label{CS2}
\end{eqnarray}
Hence, the three Casimir invariants $c_{i}$ ($i=1,2,3$) of the conformal
algebra may be built on these quantities
\begin{eqnarray}
\left( C_{\mu },c_{i}\right) =0 &\qquad &c_{1}=\hbar V\cdot Q-S^{2}
\nonumber \\
c_{2}=S\cdot Q &\qquad &c_{3}=\frac{Q^{2}}{2}-S^{2}  \label{cas}
\end{eqnarray}

Casimir invariants may then be used to reduce the quadrupole vector $%
Q_{\mu }$ as a sum of terms lying along velocity and spin vectors and of an
extra transverse part $\widehat{Q}_{\mu }$
\begin{eqnarray}
&&Q_{\mu }=\left( V\cdot Q\right) V_{\mu }+\alpha S_{\mu }+\widehat{Q}_{\mu }
\nonumber \\
&&V\cdot Q=\frac{c_{1}}{\hbar }-\hbar s\left( s+1\right)   \nonumber \\
&&\alpha =-\frac{c_{2}}{\hbar ^{2}s\left( s+1\right) }  \label{defQh}
\end{eqnarray}
Commutators between vectors $R_{\mu }$ and $\widehat{Q}_{\mu }$ may be
written
\begin{eqnarray}
&&\left( \widehat{Q}_{\mu },\widehat{Q}_{\nu }\right) =\beta S_{\mu \nu
}\qquad \left( R_{\mu },R_{\nu }\right) =\gamma S_{\mu \nu }  \nonumber \\
&&\left( R_{\mu },\widehat{Q}_{\nu }\right) =\hbar \gamma \left( \eta _{\mu
\nu }-V_{\mu }V_{\nu }\right) -\frac{\widehat{Q}_{\mu }\cdot \widehat{Q}%
_{\nu }}{\hbar }+\beta \frac{S_{\mu }\cdot S_{\nu }}{\hbar }  \nonumber \\
&&\beta =2\frac{V\cdot Q}{\hbar }-\left( \frac{S\cdot Q}{S^{2}}\right)
^{2}\qquad \gamma =\frac{\widehat{Q}^{2}-\beta S^{2}}{\hbar ^{2}}
\label{QhR}
\end{eqnarray}
Since the coefficients $V\cdot Q$, $\alpha $, $\beta $ and $\gamma $ may be
expressed in terms of Casimir invariants (\ref{cas}) and of spin number $s$,
they commute with Poincar\'{e} and dilatation generators and with each
other. However, they do not commute with $R_{\mu }$, $\widehat{Q}_{\mu }$, $%
Q_{\mu }$ or $C_{\mu }$.

As shown by relation (\ref{defR}), only the transverse part $\widehat{Q}%
_{\mu }$ of quadrupole momenta is involved in the variation of squared spin
or in the definition of $R_{\mu }$. We may therefore express the condition
of invariance of the squared spin $S^{2}$ or of the spin number $s$ as the
vanishing of $R_{\mu }$ or equivalently of $\widehat{Q}_{\mu }$. In the case
of an arbitrary $2-$photon state, relations (\ref{gconf2}) show that $Q_{\mu
}$ only contains terms lying along velocity and spin vectors. Therefore, $%
\widehat{Q}_{\mu }$ and $R_{\mu }$ vanish for such states which thus
correspond to a spin number preserved under transformations to accelerated
frames. As a consequence of (\ref{CS2}), all the scalars $V\cdot Q$, $\alpha
$, $\beta $ and $\gamma $ are preserved when $S^{2}$ is preserved.
Furthermore, commutation relations (\ref{QhR}) show that $\beta $ and $%
\gamma $ vanish in this case so that $V\cdot Q$ and $\alpha $ are directly
related to each other. Then, the shifts (\ref{CX},\ref{CS}) of position and
spin observables are greatly simplified since the terms proportional to
transverse quadrupoles $\widehat{Q}$ vanish. Even in this simple case
however, there remain corrections associated with quadrupoles components
lying along velocity and spin vectors. These corrections are already present
in spinless quantum field theory in a two-dimensional space-time \cite
{PRL96,PLA96}.

\section{Step operators for the spin number}

We consider now the general case where the spin number $s$ varies under
transformations to accelerated frames. Since it has a discrete spectrum with
only integer or half integer values, its variation implies that $s$ is an
operator with an infinite spectrum rather than a pure classical number. This
operator changes under transformations to accelerated frames although its
spectrum remains the same. We show in this section how these properties
manage to remain compatible. To this aim, we first clarify the role played
by the quadrupole momenta with respect to the transformation of localisation
observables. We introduce polarisation vectors which are orthogonal to
velocity and spin vectors and obey a new kind of non commutative calculus.
Using this calculus, we finally define step operators which respectively
increment and decrement spin number $s$ along the ladders corresponding
either to integer or to half integer values.

As $R_{\mu }$, the vector $\widehat{Q}_{\mu }$ is a weightless vector
orthogonal to velocity and spin and it obeys (\ref{transR}) with $R_{\mu }$
replaced by $\widehat{Q}_{\mu }$. Commutators of $\widehat{Q}_{\mu }$ with
spin are given by (\ref{XVector}) with the transversality
simplification. Hence, the following operator
vanishes when applied onto vectors $R_{\rho }$ and $\widehat{Q}_{\rho }$
\begin{eqnarray}
-S^{2}V_{\mu }V^{\rho } &-&S_{\mu }S^{\rho }=S_{\mu }{}^{\nu }S_{\nu
}{}^{\rho }-i\hbar S_{\mu }{}^{\rho }+\hbar ^{2}s\left( s+1\right) \eta
_{\mu }^{\rho }  \nonumber \\
&=&\left( S_{\mu }{}^{\nu }+i\hbar s\eta _{\mu }^{\nu }\right) \left( S_{\nu
}{}^{\rho }-i\hbar \left( s+1\right) \eta _{\nu }^{\rho }\right)
\end{eqnarray}
This is also the case for any vector obtained as a linear superposition of $%
R_{\rho }$ and $\widehat{Q}_{\rho }$ with coefficients which may depend on
the spin number $s$. These vectors constitute a linear space which we will
call the polarisation space and consists in all transverse quadrupoles
compatible with given velocity and spin vectors. The two vectors $R_{\mu }$
and $\widehat{Q}_{\mu }$ are orthogonal in the polarisation space. Their
symmetrised scalar product vanishes and their vectorial product is aligned
along $S_{\mu \nu }$
\begin{eqnarray}
R\cdot \widehat{Q} &=&0  \nonumber \\
R_{\mu }\widehat{Q}_{\nu }-R_{\nu }\widehat{Q}_{\mu } &=&\widehat{Q}_{\nu
}R_{\mu }-\widehat{Q}_{\mu }R_{\nu }=-\frac{\widehat{Q}^{2}}{\hbar }S_{\mu
\nu }  \label{prodQR}
\end{eqnarray}

In the polarisation space, a multiplication by $S_{\mu }{}^{\nu }$ appears
as a rotation operator. This geometrical picture must be dealt with
carefully since coefficients depending on $s$ do not commute with the basis
vectors $R_{\mu }$ and $\widehat{Q}_{\mu }$, while these vectors do not
commute with the spin vector
\begin{eqnarray}
S_{\mu }{}^{\rho }\cdot R_{\rho } &=&\left( S_{\mu }{}^{\rho }-i\hbar \eta
_{\mu }^{\rho }\right) R_{\rho }=R_{\rho }\left( S_{\mu }{}^{\rho }+i\hbar
\eta _{\mu }^{\rho }\right)   \nonumber \\
S_{\mu }{}^{\rho }\cdot \widehat{Q}_{\rho } &=&\left( S_{\mu }{}^{\rho
}-i\hbar \eta _{\mu }^{\rho }\right) \widehat{Q}_{\rho }=\widehat{Q}_{\rho
}\left( S_{\mu }{}^{\rho }+i\hbar \eta _{\mu }^{\rho }\right)
\label{rotatpol}
\end{eqnarray}
The definition (\ref{defR}) of $R_{\mu }$ may be written as such a rotation
operation and a similar relation holds for $\widehat{Q}_{\mu }$
\begin{equation}
R_{\mu }=-\frac{S_{\mu }{}^{\nu }}{\hbar }\cdot \widehat{Q}_{\nu }\qquad
\frac{S^{2}}{\hbar }\cdot \widehat{Q}_{\mu }=-S_{\mu }{}^{\nu }\cdot R_{\nu }
\end{equation}
Using these relations, we may build up superpositions of $\widehat{Q}_{\mu }$
and $R_{\mu }$ which are eigenvectors of the rotation operator
\begin{eqnarray}
\left\{ S_{\mu }{}^{\nu }+i\hbar s\eta _{\mu }^{\nu }\right\} \left\{ R_{\nu
}+is\widehat{Q}_{\nu }\right\}  &=&0  \nonumber \\
\left\{ S_{\mu }{}^{\nu }-i\hbar \left( s+1\right) \eta _{\mu }^{\nu
}\right\} \left\{ R_{\nu }-i\left( s+1\right) \widehat{Q}_{\nu }\right\}
&=&0  \label{eigen}
\end{eqnarray}
These vectors behave as eigenpolarisations of standard electromagnetic
theory but, once again, the coefficients appearing in the superpositions
depend on the spin number $s$ and do not commute with the basis vectors.
Relations (\ref{rotatpol}-\ref{eigen}) thus appear to define a non
commutative calculus in the polarisation space.

Using this calculus, we may introduce step operators which respectively
increment and decrement the spin number. Precisely, we define operators $%
A_{\mu }^{\pm }$ through the following relations
\begin{eqnarray}
\sqrt{s_{*}}\widehat{Q}_{\mu }\sqrt{s_{*}} &=&A_{\mu }^{+}+A_{\mu }^{-}
\nonumber \\
\sqrt{s_{*}}R_{\mu }\sqrt{s_{*}} &=&is_{*}\cdot \left( A_{\mu }^{+}-A_{\mu
}^{-}\right)   \nonumber \\
s_{*} &=&s+\frac{1}{2}  \label{QRApm}
\end{eqnarray}
We have used a new representation $s_{*}$ of the spin number in order to
simplify the form of forthcoming expressions. The relations (\ref{QRApm})
may conversely be written
\begin{eqnarray}
2A_{\mu }^{\mp } &=&\sqrt{s_{*}}\widehat{Q}_{\mu }\sqrt{s_{*}}\pm \frac{1}{%
\sqrt{s_{*}}}\left( \frac{\widehat{Q}_{\mu }}{2}+iR_{\mu }\right) \sqrt{s_{*}%
}  \nonumber \\
&=&\sqrt{s_{*}}\widehat{Q}_{\mu }\sqrt{s_{*}}\mp \sqrt{s_{*}}\left( \frac{%
\widehat{Q}_{\mu }}{2}-iR_{\mu }\right) \frac{1}{\sqrt{s_{*}}}
\label{defApm}
\end{eqnarray}
The operators $A_{\mu }^{\pm }$ are eigenpolarisations as in (\ref{eigen})
\begin{eqnarray}
S_{\mu }{}^{\nu }A_{\nu }^{\mp } &=&i\hbar \left( \frac{1}{2}\pm
s_{*}\right) A_{\mu }^{\mp }  \nonumber \\
A_{\nu }^{\mp }S_{\mu }{}^{\nu } &=&i\hbar A_{\mu }^{\mp }\left( -\frac{1}{2}%
\pm s_{*}\right)
\end{eqnarray}
They are transverse vectors obeying (\ref{XVector}) and, at the same
time, step operators which respectively increment or decrement the spin
number by unity
\begin{equation}
A_{\mu }^{\pm }s_{*}=\left( s_{*}\pm 1\right) A_{\mu }^{\pm }  \label{step}
\end{equation}

Components of incrementing operators commute as well as components of
decrementing operators while incrementing and decrementing components do not
\begin{eqnarray}
\left( A_{\mu }^{+},A_{\nu }^{+}\right) &=&\left( A_{\mu }^{-},A_{\nu
}^{-}\right) =0  \nonumber \\
\left( A_{\mu }^{+},A_{\nu }^{-}\right) &=&-\frac{i\hbar }{2}\gamma
s_{*}\left( \eta _{\mu \nu }-V_{\mu }V_{\nu }\right)  \nonumber \\
&+&\frac{1}{2}\left( \gamma -\frac{\beta }{4}\right) S_{\mu \nu }-\frac{i}{2}%
\beta s_{*}\frac{S_{\mu }\cdot S_{\nu }}{\hbar }  \label{ApAm}
\end{eqnarray}
Here again, these relations are reminiscent of commutation relations of
annihilation and creation operators of standard electromagnetic theory with
however richer properties. As a matter of fact, these commutators are
functions of Poincar\'{e} generators and scalars rather than pure numbers.

Incrementing and decrementing operators could have been defined differently,
for instance by multiplying $A_{\mu }^{\pm }$ by arbitrary functions of the
spin number. The commutators in the first line of (\ref{ApAm}) would thus
remain unchanged. Meanwhile the commutators in the second line could no
longer be written in terms of Poincar\'{e} generators only and they would
contain for example terms proportional to $\widehat{Q}_{\mu }\cdot \widehat{Q%
}_{\nu }$. This is precisely the reason why we have chosen the definition (%
\ref{defApm}).

Other remarkable relations are obtained for some tensor and scalar
expressions defined as quadratic forms of the step operators
\begin{eqnarray}
&&A_{\mu }^{\pm }A_{\nu }^{\mp }-A_{\nu }^{\pm }A_{\mu }^{\mp }=\frac{i\hbar
}{2}\left( 1\pm s_{*}\right) \left( \gamma -\beta \left( \frac{1}{2}\mp
s_{*}\right) ^{2}\right) S_{\mu \nu }  \nonumber \\
&&A_{\mu }^{\pm }A^{\mp \ \mu }=\frac{\hbar ^{2}}{2}\left( 1\pm s_{*}\right)
\left( \frac{1}{2}\pm s_{*}\right) \left( \gamma -\beta \left( \frac{1}{2}%
\mp s_{*}\right) ^{2}\right)   \nonumber \\
&&A_{\mu }^{\pm }A^{\pm \ \mu }=0  \label{AA}
\end{eqnarray}

Notice that the squared spin $S^{2}$ is unchanged when the sign of $s_{*}$
is changed. This means that negative values of the spin number $s_{*}$ may
be chosen as well as positive ones. Relations (\ref{AA}) as other ones
previously written in this section are preserved when $A_{\mu }^{+}$ and $%
A_{\mu }^{-}$ are substituted to each other while the sign of $s_{*}$ is
changed. This symmetry indicates that negative values of $s_{*}$ play the
same role as positive ones. The step operators $A_{\mu }^{\pm }$ increment
and decrement the spin numbers along ladders corresponding respectively to
integer and half-integer values of $s$, that is also half-integer and
integer values of $s_{*}$. Expressions (\ref{AA}) vanish for the particular
spin numbers $s_{*}=\pm \frac{1}{2}$ and $s_{*}=\pm 1$ which correspond to
the fundamental rungs of the ladders.

\section{Summary}

In this paper, we have confronted the physical requirements associated with
a relativistic conception of localisation in space-time with those arising
from quantum theory. In close connection with Einstein's conception of
synchronisation or localisation through the exchange of electromagnetic
pulses, we have built up our derivations upon the conformal algebra which
expresses the symmetries of electromagnetic theory.

We have given a complete definition of the observables of interest for this
problem, namely position, spin and quadrupole observables. We have also
described their shifts under frame transformations, including the case of
accelerated frames, and shown that these shifts may be written in terms of
the same observables. We have found that the redshift of mass naturally fits
the equivalence principle whereas the shifts of other localisation
observables under transformations to accelerated frames differ from
predictions of classical relativity.

Collecting the results of equations (\ref{defQ},\ref{defQh}), we
obtain the final expression of the generators of transformations to
accelerated frames
\begin{eqnarray}
C_{\mu } &=&2D\cdot X_{\mu }-P_{\mu }\cdot X^{2}+2X^{\rho }\cdot S_{\rho \mu
}-\frac{P_{\mu }}{M^{2}}S^{2}  \nonumber \\
&&+\frac{\hbar }{M}\left( \hbar \left( \alpha ^{2}+\beta \right) V_{\mu
}+2\alpha S_{\mu }+2\widehat{Q}_{\mu }\right)  \nonumber \\
&&\alpha ^{2}+\beta =2\left( \frac{c_{1}}{\hbar ^{2}}-s\left( s+1\right)
\right)  \nonumber \\
&&\alpha =-\frac{c_{2}}{\hbar ^{2}s\left( s+1\right) }
\end{eqnarray}
Proceeding similarly with (\ref{CX},\ref{CS},\ref{defQh}), we write the
shifts of position and spin observables as
\begin{eqnarray}
\left( \Delta _{a},X_{\nu }\right) &=&\frac{a_{\nu }}{2}X^{2}-\left( a\cdot
X\right) \cdot X_{\nu }  \nonumber \\
&&+\frac{\left( a\cdot S\right) \cdot S_{\nu }}{M^{2}}-\frac{a_{\nu }}{2}%
\frac{S^{2}}{M^{2}}  \nonumber \\
&&+\frac{\hbar ^{2}}{M^{2}}\left( \alpha ^{2}+\beta \right) \left( \frac{%
a_{\nu }}{2}-a\cdot VV_{\nu }\right)  \nonumber \\
&&-\frac{\hbar \alpha }{M^{2}}\left( a\cdot SV_{\nu }+a\cdot VS_{\nu }\right)
\nonumber \\
&&-\frac{\hbar }{M^{2}}\left( a\cdot \widehat{Q}V_{\nu }+a\cdot V\widehat{Q}%
_{\nu }\right)  \nonumber \\
\left( \Delta _{a},S_{\nu }\right) &=&a_{\nu }X\cdot S-a^{\mu }S_{\mu }\cdot
X_{\nu }  \nonumber \\
&&+\frac{\hbar a^{\mu }}{M}\left( \alpha S_{\mu \nu }+\widehat{Q}_{\mu \nu
}\right)  \label{final}
\end{eqnarray}
Only the contributions proportional to positions would have been obtained in
classical relativity. All the other terms may be considered as quantum
corrections associated either with spin or with quadrupole observables.

We have emphasised a particularly important result which concerns spin
transformation. The existence of transverse quadrupole corrections leads to
a variation of the spin number under transformations to accelerated frames.
It is only in the peculiar case when these corrections vanish that the spin
number may be considered as a classical number, as it is usual in standard
quantum field theory. This occurs for example when the localisation
procedure is performed with $2-$photon states. In the general case in
contrast, transverse quadrupoles do not vanish, so that the spin number has
to be treated as an operator. Its spectrum is an infinite ladder
corresponding either to integer or to half-integer values. It remains
unchanged under transformations to accelerated frames whereas the various
eigenvectors are mixed. We have characterised these transformations through
the introduction of a non commutative calculus in a polarisation space
orthogonal to velocity and spin. The shift of spin number is thus determined
by step operators which increment or decrement $s$ along the ladders of spin
eigenvalues.

\section{Prospects}

These results clearly challenge the commonly used theoretical methods where
quantum and relativistic aspects are dealt with by combining quantum field
theory on one side and classical relativity on the other one.

Quantum corrections appearing in equations (\ref{final}) are proportional to
spin or quadrupole observables and they have their orders of magnitude
essentially determined by a single length scale $\frac{\hbar }{M}$. Clearly,
they have to be interpreted as resulting from irreducible size effects
arising from the quantum nature of observables. It is therefore natural that
difficulties are met when trying to represent relativistic effects by
transformations described by infinitesimal differential geometry and acting
on sizeless points. In contrast, the results obtained in the present paper
rely on quantum algebraic techniques embedding the symmetries of
relativistic space-time and are thence more reliable than those based upon a
classical representation of space-time.

As often emphasised, the results obtained in the present paper have been
derived from conformal symmetry of electromagnetism. It is nevertheless hard
to refrain from thinking that they are worth of consideration in a more
general theoretical context. If we consider for example an annihilation
process where an electron and a positron are transformed into $2$ photons,
the position in space-time of the $2-$photon coincidence event has to be
identified as the position in space-time of the annihilation event. As
explained in appendix \ref{appLocalisation}, this position is just $X_{\mu }$
in the specific case of a $2-$photon state. This means that the position of
a physical event involving electrons has been defined.

The case of a $2-$photon state corresponds to the particular situation where
the transverse quadrupoles vanish. Hence, the spin number may still be used
as a classical number characterising an elementary representation of quantum
field theory while the shifts of observables are given by simpler relations (%
\ref{final}) with $\widehat{Q}$ and $\beta $ set to zero. But there also
exist composite quantum systems, such as atoms for example, for which there
is no fundamental reason for transverse quadrupoles to vanish. Then,
transformations to accelerated frames can no longer be described with the
finite dimensional representations of standard quantum field theory. A
consistent description must involve the full content of conformal algebra
and this unavoidably leads to infinite dimensional representations where the
different eigenvalues of spin lying along an infinite ladder have to be
simultaneously dealt with. These new features will have to be taken into
account, at some level of accuracy, when analysing experiments where atoms
are placed in acceleration fields \cite{Borde97,YoungKC97}.

On the metrological side, it has to be emphasised that the definition of
units is more and more evolving towards the use of quantum standards. This
evolution not only results of technological progress but, more basically, of
efforts to improve the universality of the definition of units. Dilatation
symmetry plays a central role in this context as soon as dilatation is
understood as a correlated change of time, space and mass scales which
preserves the velocity of light and the Planck constant \cite
{Dicke62,Sakharov74,Hoyle75}. An appropriate behaviour under dilatations is
needed to ensure universality of the relations which connect the electron
mass to its Compton length or to the Rydberg constant. In the present paper,
we have shown that mass defined as a Lorentz scalar for a field state varies
according to the change of the conformal factor under dilatations or
transformations to accelerated frames. This is just the expression of the
equivalence principle or, equivalently in a metrological context, of the
universality of the definition of units. Obviously, metrological definitions
not only rely on the physics of electromagnetic fields but also on the
physics of atoms and electrons. Hence, these metrological reflections appeal
for an enlargement of the present theory of electrons which should
incorporate a more complete implementation of symmetries within the algebra
of quantum observables.

\appendix

\section{Conformal invariance of the photon number}

\label{appConfInv}

We briefly discuss in this appendix the explicit realisation of the
conformal algebra with quantum fields. As its practical representation will
be given by the propagating fields used when performing time transfer and
localisation in space-time, we shall be concerned with free fields only.
Within the context of Quantum Field Theory, the generators of propagation
symmetries can be constructed as integrals of the energy-momentum tensor of
the field, that is also as quadratic forms of the quantum fields. Explicit
expressions may be found for instance in \cite{ItzyksonZ85}. However, these
expressions will not be needed in the following, which will only use the
general transformation properties of fields under these symmetries.

When written with normally ordered products, the generators $\Delta$
are found to
vanish in the vacuum state $\left| {\rm {vac}}\right\rangle $
\begin{equation}
\Delta \left| {\rm {vac}}\right\rangle =0  \label{vacinv}
\end{equation}
Such a property is made consistent by the conformal invariance of
electromagnetic vacuum \cite{QSO95}. More generally the definition of the
number of photons is also conformally invariant in electromagnetic theory
\cite{Gross64}. This property may be written by introducing the projector $%
\Pi _{n}$ on the space of $n-$photon states
\begin{equation}
\left( \Delta ,\Pi _{n}\right) =0  \label{phonuminv}
\end{equation}

Consider now the generic $1-$photon state built through the action of an
arbitrary field operator on vacuum
\begin{equation}
\phi ^{\dagger }=\sum_{i}\varphi _{i}a_{i}^{\dagger }  \label{deffield}
\end{equation}
$\phi ^{\dagger }$ is in fact the negative frequency part of a field, that
is also an arbitrary linear superposition of creation operators $%
a_{i}^{\dagger }$ where $i$ completely characterises the field modes, for
instance by their momentum and polarisation, and $\varphi _{i}$ are
classical field amplitudes. Due to (\ref{phonuminv}), the action of a
generator $\Delta $ on this $1-$photon state is another $1-$photon state. It
follows from conformal invariance (\ref{vacinv}) of vacuum that this state
may be expressed as
\begin{eqnarray}
\Delta \phi ^{\dagger }\left| {\rm vac}\right\rangle  &=&\left( \Delta
\phi ^{\dagger
}-\phi ^{\dagger }\Delta \right) \left| {\rm vac}\right\rangle   \nonumber \\
&=&i\hbar \left( \Delta ,\phi ^{\dagger }\right) \left| {\rm vac}\right\rangle
\label{comp1}
\end{eqnarray}
$\left( \Delta ,\phi ^{\dagger }\right) $ is a linear superposition of creation
operators like $\phi ^{\dagger }$ which, therefore, commutes with $\phi
^{\dagger }$ as well as with other expressions $\left( \Delta ^{\prime },\phi
^{\dagger }\right) $ of the same kind. The product of operators in the
algebra then translates into the composition of their commutators
\begin{eqnarray}
\Delta \Delta ^{\prime }\phi ^{\dagger }\left| {\rm vac}\right\rangle
  &=&i\hbar \Delta \left(
\Delta ^{\prime },\phi ^{\dagger }\right)
 \left| {\rm vac}\right\rangle   \nonumber
\\
&=&-\hbar ^{2}\left( \Delta ,\left( \Delta ^{\prime },\phi ^{\dagger }\right)
 \right)
\left| {\rm vac}\right\rangle   \label{comp}
\end{eqnarray}

Fields and energy-momentum operators do not commute in general since
propagating fields are not invariant under translation, but the following
relation results from the massless character of the electromagnetic field
implied by Maxwell equations
\begin{equation}
\left( P^{\mu },\left( P_{\mu },\phi ^{\dagger }\right) \right) =0
\label{m1}
\end{equation}
The vanishing mass of $1-$photon states is then seen to result from
relations (\ref{comp}) and (\ref{m1}). Precisely, one demonstrates the
following equivalent relations
\begin{eqnarray}
P^{2}\phi ^{\dagger }\left| {\rm vac}\right\rangle &=&0  \nonumber \\
P^{2}\Pi _{1} &=&0  \label{mass}
\end{eqnarray}

In the same manner, $2-$photon states can be built as the result $\phi
_{1}^{\dagger }\phi _{2}^{\dagger }\left| {\rm vac}\right\rangle $ of the
action of two field operators defined as in (\ref{deffield}) on vacuum. The
actions of generator $\Delta $ on these states are other $2-$photon states
obtained through the following relations which have to be compared with
relations (\ref{comp1}) holding for $1-$photon states
\begin{eqnarray}
&&\Delta \phi _{1}^{\dagger }\phi _{2}^{\dagger }\left| {\rm vac}\right\rangle
=i\hbar \left( \Delta ,\phi _{1}^{\dagger }\phi _{2}^{\dagger }\right)
 \left| {\rm %
vac}\right\rangle   \nonumber \\
&&\left( \Delta ,\phi _{1}^{\dagger }\phi _{2}^{\dagger }\right)
=\left( \Delta ,\phi
_{1}^{\dagger }\right) \phi _{2}^{\dagger }+\phi _{1}^{\dagger }\left(
\Delta ,\phi _{2}^{\dagger }\right)   \label{gen2}
\end{eqnarray}
One proceeds similarly for describing the action of two generators $\Delta $
 and $%
\Delta ^{\prime }$ on the same $2-$photon state
\begin{eqnarray}
\Delta \Delta ^{\prime }\phi _{1}^{\dagger }\phi _{2}^{\dagger }\left|
 {\rm vac}%
\right\rangle  &=&-\hbar ^{2}\left( \Delta ,
\left( \Delta ^{\prime },\phi _{1}^{\dagger
}\phi _{2}^{\dagger }\right) \right) \left| {\rm vac}\right\rangle
\nonumber \\
\left( \Delta ,\left( \Delta ^{\prime },\phi _{1}^{\dagger }
\phi _{2}^{\dagger }\right)
\right)  &=&\left( \Delta ,\left( \Delta ^{\prime },
\phi _{1}^{\dagger }\right) \right)
\phi _{2}^{\dagger }  \nonumber \\
&&+\left( \Delta ,\phi _{1}^{\dagger }\right) \left( \Delta ^{\prime },\phi
_{2}^{\dagger }\right)   \nonumber \\
&&+\left( \Delta ,\phi _{2}^{\dagger }\right) \left( \Delta ^{\prime },\phi
_{1}^{\dagger }\right)   \nonumber \\
&&+\phi _{1}^{\dagger }\left( \Delta ,\left( \Delta ^{\prime },
\phi _{2}^{\dagger
}\right) \right)   \label{comp2}
\end{eqnarray}
Notice that the product of actions on different fields is commutative.
According to relation (\ref{gen2}), the symmetry generators can be
decomposed as sums of actions on a single field
\begin{eqnarray}
&&\Delta \Pi _{2}=\left( \Delta ^{(1)}+\Delta ^{(2)}\right) \Pi _{2}
  \nonumber \\
&&\left( \Delta ^{(1)},\phi _{1}^{\dagger }\phi _{2}^{\dagger }\right) =\left(
\Delta ,\phi _{1}^{\dagger }\right) \phi _{2}^{\dagger }  \nonumber \\
&&\left( \Delta ^{(2)},\phi _{1}^{\dagger }\phi _{2}^{\dagger }\right) =\phi
_{1}^{\dagger }\left( \Delta ,\phi _{2}^{\dagger }\right)   \label{decomp}
\end{eqnarray}
Relation (\ref{comp2}) may then be understood as exhibiting the distributive
property of the product of operators
\begin{equation}
\Delta \Delta ^{\prime }\Pi _{2}=\left( \Delta ^{(1)}+
\Delta ^{(2)}\right) \left( {\Delta ^{\prime}}^{(1)}
+{\Delta ^{\prime}}^{(2)}\right) \Pi _{2}
\end{equation}
Symmetry generators acting on single fields furthermore satisfy equation (%
\ref{mass}).

\section{Synchronisation with one-photon states}

\label{appSynchronisation}

The quantum description of time transfer has been described in detail using
the simple model of scalar field theory in two-dimensional (2d) space-time
\cite{PRL96,BJP95}. This description heavily relied on a specific feature of
2d quantum field theories, namely the existence of an {\it a priori}
decomposition of fields in counterpropagating directions. In the present
appendix, we develop a quantum description of time transfer performed in
four-dimensional (4d) space-time by using electromagnetic $1-$photon states.

We start from relations (\ref{m1}) and (\ref{mass}) which result from the
massless character of the electromagnetic field. A whole set of other
relations results from the conformal invariance of Maxwell equations \cite
{Bateman09,Cunningham09,BinegarFH83}. Transforming (\ref{m1}) and (\ref{mass}%
) under the action of conformal generators (\ref{ConfAlg}), one obtains
\begin{eqnarray}
0 &=&P^{2}\Pi _{1}  \nonumber \\
0 &=&\left( P^{\lambda }\cdot J_{\lambda \mu }+P_{\mu }\cdot D\right) \Pi
_{1}  \nonumber \\
0 &=&\left( 2J_{\ \mu }^{\lambda }\cdot J_{\lambda \nu }+P_{\mu }\cdot
C_{\nu }+P_{\nu }\cdot C_{\mu }\right) \Pi _{1}  \nonumber \\
&&+\eta _{\mu \nu }\left( 2D^{2}-P\cdot C\right) \Pi _{1}  \nonumber \\
0 &=&\left( C^{\lambda }\cdot J_{\lambda \mu }-C_{\mu }\cdot D\right) \Pi
_{1}  \nonumber \\
0 &=&C^{2}\Pi _{1}  \label{mconf}
\end{eqnarray}

When taken together, relations (\ref{mconf}) constitute a conformal
invariant characterisation of $1-$photon states which has interesting
consequences. The first two of these relations entail that spin, as defined
by Pauli-Lubanski vector $W_{\mu }$ (\ref{PW}), is proportional to momentum
for $1-$photon states
\begin{eqnarray}
-\left( P_{\lambda }J_{\mu \nu }+P_{\mu }J_{\nu \lambda }+P_{\nu }J_{\lambda
\mu }\right) \Pi _{1} &=&\epsilon _{\lambda \mu \nu \rho }W^{\rho }\Pi _{1}
\nonumber \\
&=&\sigma \epsilon _{\lambda \mu \nu \rho }P^{\rho }\Pi _{1}  \label{hel}
\end{eqnarray}
$\sigma $ is a Casimir invariant of the whole conformal algebra. In fact,
relations (\ref{mconf}) allow one to deduce the three Casimir invariants of
the conformal algebra from $\sigma $ for $1-$photon states.

Then, transfer variables $U_{\mu }$ can be associated with a given $1-$
photon state. These transfer variables are defined so that the Poincar\'{e}
and dilatation generators have their classical form
\begin{eqnarray}
J_{\mu \nu }\Pi _{1} &=&\left( P_{\mu }\cdot U_{\nu }-P_{\nu }\cdot U_{\mu
}+S_{\mu \nu }\right) \Pi _{1}  \nonumber \\
D\Pi _{1} &=&P_{\mu }\cdot U^{\mu }\Pi _{1}  \label{transSync}
\end{eqnarray}
As a consequence of the vanishing mass, the transfer variables $U_{\mu }$
are not uniquely defined by relations (\ref{transSync}). They characterise
the position of the photon transversely to propagation but their
longitudinal components are not defined. This is not a defect but on the
contrary a necessary feature for transfer observables used to exchange
information between two remote observers.
Using (\ref{hel}), one may for instance define
transfer variables as
\begin{equation}
U_{\mu }={\frac{1}{P_{0}}}\cdot J_{0\mu }
\end{equation}
This definition can be seen as generalising the time transfer variables
defined in a 2d quantum field theory \cite{PRL96} to four dimensional
space-time.

Then, the third relation in (\ref{mconf}) can be used to solve for the
generators of transformations to accelerated frames in terms of Poincar\'{e}
and dilatation generators and the Casimir invariant $\sigma $. Using (\ref
{mconf},\ref{transSync}), it is then possible to rewrite conformal
generators in terms of the transfer variables and to deduce the shifts of
transfer observables under transformations to accelerated frames, thus
generalising expressions known for 2d quantum field theory \cite{PRL96}.

\section{Localisation with two-photon states}

\label{appLocalisation}

We now proceed similarly for the problem of localisation. As explained in
detail in \cite{PLA96,EPL97}, the definition of a localised event requires
the presence of two photons propagating in different directions. The
corresponding quantum state thus corresponds to a non vanishing mass.

We first evaluate the mass associated with the $2-$photon state, using the
decomposition (\ref{decomp}) of symmetry generators $\Delta $ on operators $%
\Delta ^{(1)} $ and $\Delta ^{(2)}$ acting on each field.
We also use the algebraic
relations (\ref{mass}) associated with massless fields for symmetry
generators $\Delta ^{(1)}$ and $\Delta ^{(2)}$
as well as their transformed relations (%
\ref{mconf}) under conformal symmetry. In particular, the momentum of the $%
2- $photon state is the sum of two momenta each corresponding to a vanishing
mass so that the resulting mass is obtained as the product of these momenta
\begin{eqnarray}
P^{2}\Pi _{2} &=&\left( P^{(1)}+P^{(2)}\right) ^{2}\Pi _{2}  \nonumber \\
&=&2P^{(1)\mu}P^{(2)}_\mu \Pi _{2}
\end{eqnarray}
This mass does not vanish for $2-$photon states with non parallel momenta.
As a consequence, positions $X_{\mu }$ describing localisation in space-time
can be defined from the symmetry generators according to the general
definition (\ref{defX}).

In the particular case of $2-$photon states, we may give a geometrical
interpretation of the definition of $X_{\mu }$ through the following
argument. We first introduce space-time variables ${X_{\mu }^{(1)}}$ and ${%
X_{\mu }^{(2)}}$ for each of the two photons
\begin{eqnarray}
{\frac{1}{2}}P^{2}\cdot {X_{\mu }^{(1)}}\Pi _{2} &=&\left( P^{\lambda }\cdot
{J}_{\lambda \mu }^{(1)}+P{_{\mu }^{(1)}}\cdot D\right) \Pi _{2}  \nonumber
\\
{\frac{1}{2}}P^{2}\cdot {X_{\mu }^{(2)}}\Pi _{2} &=&\left( P^{\lambda }\cdot
{J}_{\lambda \mu }^{(2)}+P{_{\mu }^{(2)}}\cdot D\right) \Pi _{2}
\label{trans2}
\end{eqnarray}
These space-time variables correspond to particular choices of the transfer
variables $U_{\mu }$ introduced for $1-$photon states through relations (\ref
{transSync}). The total momentum of the $2-$ photon state has been used to
raise the ambiguity on the longitudinal component of these variables in a
Lorentz covariant way. Then, the space-time position of the $2-$photon
system corresponds to half the sum of these two variables
\begin{equation}
X_{\mu }\Pi _{2}={\frac{{X_{\mu }^{(1)}}+{X_{\mu }^{(2)}}}{2}}\Pi _{2}
\label{position2}
\end{equation}
In a classical approximation, a $1-$photon state may be represented as a
light pulse and a $2-$photon state by two light pulses \cite{PLA96}. In a 2d
theory, two counterpropagating light pulses have to meet at some space-time
position which is just the position $X_{\mu }$. In a 4d theory in contrast,
two rays do not necessarily meet each other but the relations (\ref{trans2}-%
\ref{position2}) nevertheless provide a generalised geometrical
interpretation. If two rays $r^{(1)}$ and $r^{(2)}$ represent the
trajectories of the two photons in space-time and if $r^{\perp }$ is defined
as the straight line which crosses these two rays at right angle, then ${%
X_{\mu }^{(1)}}$ and ${X_{\mu }^{(2)}}$ are the intersection points of $%
r^{(1)}$ and $r^{(2)}$ with $r^{\perp }$ and ${X}_{\mu }$ is the middle
point of the segment joining ${X_{\mu }^{(1)}}$ and ${X_{\mu }^{(2)}}$.

The conformal generators acting on each photon are then deduced from
relations (\ref{mconf})
\begin{eqnarray}
{J_{\mu \nu }^{(1)}}\Pi _{2} &=&\left( P{_{\mu }^{(1)}}\cdot {X_{\nu }^{(1)}}%
-P{_{\nu }^{(1)}}\cdot {X_{\mu }^{(1)}}+S{_{\mu \nu }^{(1)}}\right) \Pi _{2}
\nonumber \\
S{_{\mu \nu }^{(1)}} &=&2\epsilon _{\mu \nu \rho \lambda }{P^{(1)\rho}}%
\frac{P^{\lambda }}{P^{2}}\sigma^{(1)}  \nonumber \\
{D^{(1)}}\Pi _{2} &=&P{_{\mu }^{(1)}}\cdot X^{(1)\mu}\Pi _{2}  \nonumber
\\
C{_{\mu }^{(1)}}\Pi _{2} &=&\left( 2D^{(1)}\cdot X{_{\mu }^{(1)}}-P{_{\mu
}^{(1)}}\cdot {X^{(1)}}^{2}+X^{(1)\lambda}\cdot {S_{\lambda \mu }^{(1)}%
}\right.   \nonumber \\
&&+\left. {\frac{P{_{\mu }^{(2)}}}{P^{2}}}\left( 4{\sigma^{(1)}}%
^{2}+1\right) \right) \Pi _{2}
\end{eqnarray}
Similar relations hold for labels $(1)$ and $(2)$ interchanged.
The sum of these
$1-$photon generators then provides an expression for symmetry generators
associated with $2-$photon states
\begin{eqnarray}
J_{\mu \nu }\Pi _{2} &=&\left( P_{\mu }\cdot X_{\nu }-P_{\nu }\cdot X_{\mu
}+S_{\mu \nu }\right) \Pi _{2}  \nonumber \\
D\Pi _{2} &=&P\cdot X\Pi _{2}  \nonumber \\
C_{\mu }\Pi _{2} &=&\left( 2D\cdot X_{\mu }-P_{\mu }\cdot X^{2}+2X^{\rho
}\cdot S_{\rho \mu }-\frac{P_{\mu }}{P^{2}}S^{2}\right.   \nonumber \\
&&\left. +\sigma ^{2}{\frac{P_{\mu }}{P^{2}}}-2\sigma {\frac{W_{\mu }}{P^{2}}%
}\right) \Pi _{2}  \label{gconf2}
\end{eqnarray}
$C_{\mu }\Pi _{2}$ is thus the sum of the external part (\ref{defCp}) of
conformal generators to accelerated frames written in terms of Poincar\'{e}
and dilatation generators and of two further terms respectively aligned
along momentum $P_{\mu }$ and Pauli-Lubanski spin vector $W_{\mu }$. This
entails that the spin number is invariant under all conformal
transformations for arbitrary $2-$photon states. The parameter $\sigma $ is
a conformal invariant of the $2-$photon state obtained by summing up the
Casimir invariants associated with each photon
\begin{equation}
\sigma =\sigma^{(1)}+\sigma^{(2)}
\end{equation}

The set of observables for the $2-$photon state may be completed by adding
to the previous ones further combinations characterising the internal
structure of the system
\begin{eqnarray}
\Delta P_{\mu } &=&P{_{\mu }^{(2)}}-P{_{\mu }^{(1)}}  \nonumber \\
\Delta X_{\mu } &=&X{_{\mu }^{(2)}}-X{_{\mu }^{(1)}}  \nonumber \\
\Delta \sigma  &=&{\sigma^{(2)}}-{\sigma^{(1)}}
\end{eqnarray}
Vectors $P$, $\Delta P$ and $\Delta X$ can be seen to describe a triad of
orthogonal vectors
\begin{eqnarray}
P\cdot \Delta P &=&P\cdot \Delta X=\Delta P\cdot \Delta X=0  \nonumber \\
\Delta P^{2} &=&-P^{2}
\end{eqnarray}
Furthermore, explicit computation shows that these quantities determine the
spin associated with the $2-$photon state
\begin{eqnarray}
W_{\mu }\Pi _{2} &=&\left( -{\frac{1}{2}}\epsilon _{\mu \nu \lambda \rho
}P^{\nu }S^{\lambda \rho }\right) \Pi _{2}  \nonumber \\
&=&\left( -{\frac{1}{2}}\epsilon _{\mu \nu \lambda \rho }P^{\nu }\Delta
P^{\lambda }\Delta X^{\rho }+\Delta P_{\mu }\Delta \sigma \right) \Pi _{2}
\end{eqnarray}
These expressions provide a simple geometric interpretation for the spin of
the $2-$photon state as the sum of two contributions. The first one is the
spatial angular momentum of the two non-intersecting rays associated with
the two photons, while the second one arises from the individual spins of
the two photons.

\end{document}